# Silicon doped graphene for improved performance of heterojunction solar cell applications

Shengjiao Zhang,[a] Shisheng Lin,*[a] Xiaoqiang Li,[a] Xiaoyi Liu,[b] Hengan Wu,[b] Peng Wang,[a] Zhiqian Wu,[a] Huikai Zhong,[a] Wenli Xu[a]  and Zhijuan Xu[a] a College of Information Science and Electronic Engineering, Zhejiang University, Hangzhou, 310027, China.

b Department of Modern Mechanics, Chinese Academy of Science Key Laboratory of Mechanical Behavior and Design of Materials, University of Science and Technology of China, Hefei, 230000, China.

* Corresponding Author. E-mail: shishenglin@zju.edu.cn

**Graphene has attracted increasing interests due to its remarkable properties, however, the zero band gap of monolayer graphene might limit its further electronic and optoelectronic applications. Herein, we have successfully synthesized monolayer silicon-doped graphene (SiG) in large area by chemical vapor deposition method. Raman spectroscopy and X-ray photoelectron spectroscopy measurements evidence silicon atoms are doped into graphene lattice with the doping level of 3.4 at%. The electrical measurement based on field effect transistor indicates that the band gap of graphene has been opened by silicon doping, which is around 0. 28 eV supported by the first-principle calculations, and the ultraviolet photoelectron spectroscopy demonstrates the work function of SiG is 0.13 eV larger than that of graphene. Moreover, the SiG/GaAs heterostructure solar cells show an improved power conversion efficiency of 33.7% in average than that of graphene/GaAs solar cells, which are attributed to the increased barrier height and improved interface quality. Our results suggest silicon doping can effectively engineer the band gap of monolayer graphene and SiG has great potential in optoelectronic device applications.**

**Introduction**

Graphene, a two-dimensional (2D) material discovered in 2004[1] with many remarkable properties, such as anomalous quantum hall effect, high electron mobility at room temperature, excellent thermal conductivity, 97.7% transmittance of visible light and so on, has attracted wide attention to various applications in electronic and optoelectronic area.[2-6] Although enormous efforts have been devoted to the research and potential applications of graphene, its intrinsic property of zero band gap casts great obstacles on its further applications. Therefore, many approaches have been taken to overcome this disadvantage.[7-11] Among such methods, doping is taken as one of the most feasible ways to open the band gap of graphene and tailor its electrical properties, for the reason that doping can break the symmetric structure of graphene. Hydrogenated graphene has been carried out to open the band gap of graphene,[12] which could reach as large as 5.4 eV via controlling the coverage and configuration of hydrogen atoms on the surface of graphene.[13] Unfortunately, the distribution and the number of hydrogen absorbed on the graphene have a significant influence on the stability of hydrogenated graphene, leading to unsustainable controlled electrical properties.[14] Compared with hydrogenation, substitutional doping will destroy the symmetric structure of graphene by replacing the hexagonal carbon atoms with dopant atoms such as N, B and Bi,[15-18] which could make the electrical properties tailored more stable in theoretical predication. In practical, N-doped graphene can be controllably realized, such as annealing in $NH_3$ atmosphere, reaction using pyridine, which would open the band gap of graphene and shift the Fermi level into conduction band.[19-22] By chemical vapor deposition (CVD) synthesis or thermal exfoliation of graphite in the atmosphere/solution comprising boron atoms, B-doped graphene can be achieved.[23-25] In particular, the band gap of graphene opened could reach about 0.7 eV via nitrogen and boron combined doping,[26,27] but the maldistribution of small BN domains will impede the application of this material in nanoscale optoelectronic devices.[28] Unlike N-doped and B-doped graphene, as an element in the same main group with carbon, silicon doping is expected to open the band gap of graphene while doesn't affect the carrier concentration. To now, silicon doped graphene (SiG) has not been well-explored both theoretically and experimentally. So far, periodic density functional theory calculations of SiG shows that the band gap will become larger with the number of dopant atoms (silicon atoms) increasing when the ratio of silicon atoms in unit cell is less than 0.5, and it will exceed 2 eV while the ratio reaches 0.5.[29] Moreover, SiG might have potential applications in gas sensors because it is more reactive when absorbing a bit of gas molecules like CO, $NO_2$, rhodamine B (RhB) and methylene blue (MB).[30-33] Compared with the occasional discovery of SiG during the CVD growth of graphene,[31] Ruitao et al. have controllably produced SiG through reaction of methoxytri-methylsilane (MTMS, $C_4H_{12}OSi$) and hexane.[33] Although SiG has been experimentally synthesized, only some characterizations of this material have been carried out and there are no electrical measurements and practical applications of this material reported in the optoelectronic field. In this paper, we found that the band gap of graphene opened by silicon doping using CVD synthesis technique experimentally is around 0.28 eV, and the application of SiG in heterostructure solar cell is investigated. The PCE of SiG/GaAs is 33.7% higher than that of the solar cell based on graphene/GaAs hetestructure in average.

**Results and discussion**

The optical image of SiG transferred to $SiO_2$/Si wafer is shown in Fig. 1a, while the inset displays the structure of SiG. From this photograph, we can see the transferred SiG is homogeneous on centimeter scale. As Raman spectra of SiG and graphene film shown in Fig. 1b, there are three main peaks assigned in the spectroscopy, i.e. D peak (1354 cm$^{-1}$), G peak (1593 cm$^{-1}$) and 2D peak (2793 cm$^{-1}$). In the Raman spectrum of graphene, the intensity ratio of 2D peak to G peak (I$_{2D}$/I$_G$=2.9) can be regarded as a typical feature of single layer graphene. Besides, the weak D peak indicates the high quality of as-synthesized graphene. Due to the absorption of water and oxygen, the position of G peak shifts to 1593 cm$^{-1}$, ndicating the graphene is p-doped.[34] No significant shifts of G peak and 2D peak can be seen in the Raman spectrum of SiG when compared with that of graphene. However, there is a strong D peak appeared in the spectrum of SiG. This phenomenon might be caused by the insertion of silicon atoms in the graphene lattice which breaks the symmetric structure of graphene[14] (the inset in Fig. 1a). Moreover, the strengthening of the intensity of D peak decreases the intensity ratio of 2D peak and G peak. Fig. 1c and Fig. 1d illustrate the XPS spectra of silicon doped and pristine graphene films. The SiG and graphene membranes measured were transferred to Ge substrate so that to avoid the influence of silicon in the Si/SiO$_2$ substrate. As shown in Fig. 1c, both SiG and graphene have a significant signal of the *sp$^2$* hybridized C-C bonds around 284.4 eV, indicating the component of graphitic carbon.[35] Besides, two signals, one peak assigned to 101.1 eV and another located at 102.8 eV, are observed during the Si2p scanning, which appear only in SiG, as shown in Fig. 1d. The former peak may relate to Si–C binding while the latter may be a symbol of Si–O–C.[36] These two peaks detected in the SiG membrane should provide a reasonable evidence for doped silicon binding to carbon atoms. Based on the further analysis of the XPS data of SiG membrane, the silicon doping level is 3.4 at%.

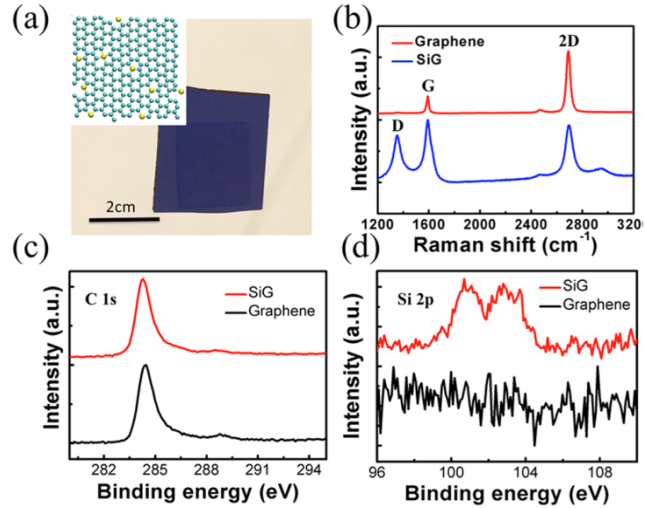

**Fig. 1** (a) The optical picture of SiG transferred to $SiO_2$/Si wafer, the inset is the schematic structure of SiG, where several carbon atoms in the lattice are replaced by silicon atoms. (b) Raman spectra of SiG and graphene on $SiO_2$/Si substrate. (c)-(d) XPS spectra of SiG and graphene sheets. As shown in (c), there is no dramatic difference of the peak of graphite-like $sp^2$ C between SiG and graphene. From (d), obvious peak of Si2p at 101.1 eV and 102.8 eV can be detected of SiG, providing a reliable evidence for silicon doping.

Fig. 2a shows the low magnification TEM image of SiG membrane. The electron diffraction image inserted in the left region of Fig. 2a shows the honeycomb structure of SiG in a nanoscale. By the aid of the folded area marked with a red rectangle, the edge of the SiG membrane was observed, illustrated in Fig. 2b. From the image, we can clearly see that the SiG film is monolayer. The thickness of SiG transferred to the $SiO_2$/Si (90 nm/500 μm) substrate was measured by AFM, which is depicted in Fig. 2c. From the image, we can see that the membrane is almost smooth in the whole area except the edge region where there are some folded and crack areas caused by the transfer process. Besides, the height of the SiG (0.37 nm) is similar to that of monolayer graphene (0.35 nm).[37]

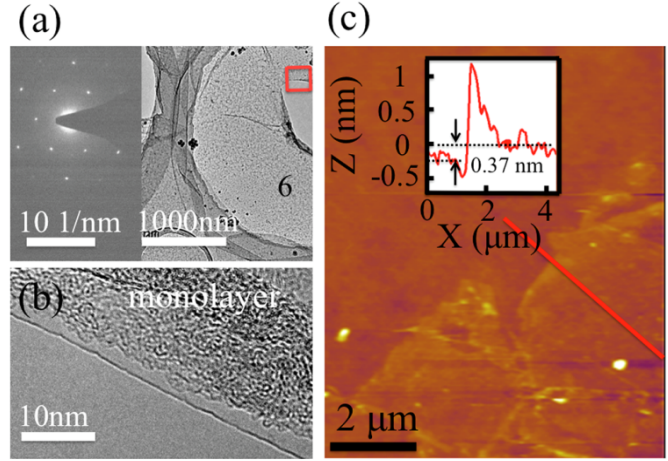

**Fig. 2** (a) Low magnification TEM image of suspended SiG membrane. The electron diffraction pattern inserted in the left of (a) shows that the structure of SiG is honeycomb. (b) HRTEM images of SiG sheet on TEM grid in the area of the red rectangle. (c) AFM image of SiG film. The image inserted demonstrates that the height of the SiG along the red line is around 0.37 nm.

The UPS of SiG and graphene are demonstrated in Fig. 3a. The kinetic energies are drawn after correcting with the applied bias of -9.8 V. The vacuum level, corresponding to the onset of the secondary electrons ($E_{sec}$), was measured by linear extrapolation of the low-kinetic energy onset (secondary electron cutoff). The work function ($\phi$) of graphene can be deduced from the equation (1):[38]

$$\varphi = h\nu - (E_{FE} - E_{sec}) \qquad (1)$$

where $h\nu=21.2$ eV (He I Source), and $E_{FE}$ is the Fermi edge of graphene. Since the $E_{FE}$ of the SiG and graphene are almost the same (See Fig. S1†), the difference of the work function between SiG and graphene depends on the value of $E_{sec}$. According to Fig. 3a, the cut-off energy ($E_{sec}$) of SiG is 0.13 eV larger than that of graphene, resulting in the work function of SiG improved by 0.13 eV compared to that of graphene according to equation. Besides, SiG and graphene were used to fabricate back-gated FET for the electrical measurement. Fig. 3b shows the structure of the FET, which includes the Au electrodes, SiG, $SiO_2$ and highly doped silicon, respectively. The SiG, bridging the source and drain electrodes, acted as the conducting channel which is 2 mm wide by 1 mm long. The transfer characteristics of current between drain and source ($I_{DS}$) vs. back gate voltage ($V_G$) are illustrated in Fig. 3c. In the measurement, the bias voltage of 1 V is fixed. According to the $I_{DS}$ vs. $V_G$ curve, the voltage at Dirac point shifts to the positive direction, originating from the absorption of oxygen and water molecules which caused the graphene to be p-doped.[34] Besides, the carrier mobility (μ) can be deduced by equation (2):

$$\mu = \frac{L}{WC_{BG}V_{DS}} * \frac{\Delta I_{DS}}{\Delta V_G} \qquad (2)$$

where $C_{BG}$ is the gate capacitance per unit area (11.5 nF·cm$^{-2}$ for 300 nm SiO$_2$),[39] L is the channel length, W is the channel wide, and $V_{DS}$ is the bias voltage. Therefore, the mobility of the FET device based on graphene is 1253 cm$^2$V$^{-1}$s$^{-1}$, in agreement with the value of CVD grown graphene reported,[40] while that of the FET based on SiC is 463 cm$^2$V$^{-1}$s$^{-1}$. In addition, compared with the $I_{DS}$ vs. $V_G$ curve of graphene, SiG has smaller electrical conductivity. All of these different electrical properties of SiG may result from the silicon dopants, which replace some carbon atoms in graphene lattice and generate defects in graphene (as indicated by Raman spectrum in Fig. 1b). These foreign atoms behaved as scattering centers, thus decreasing the mobility and the conductivity of graphene.[21] Since the silicon atoms have formed the covalent bond with carbon atoms in graphene lattice, the symmetric structure was broken, thus opening the band gap of graphene.[29]

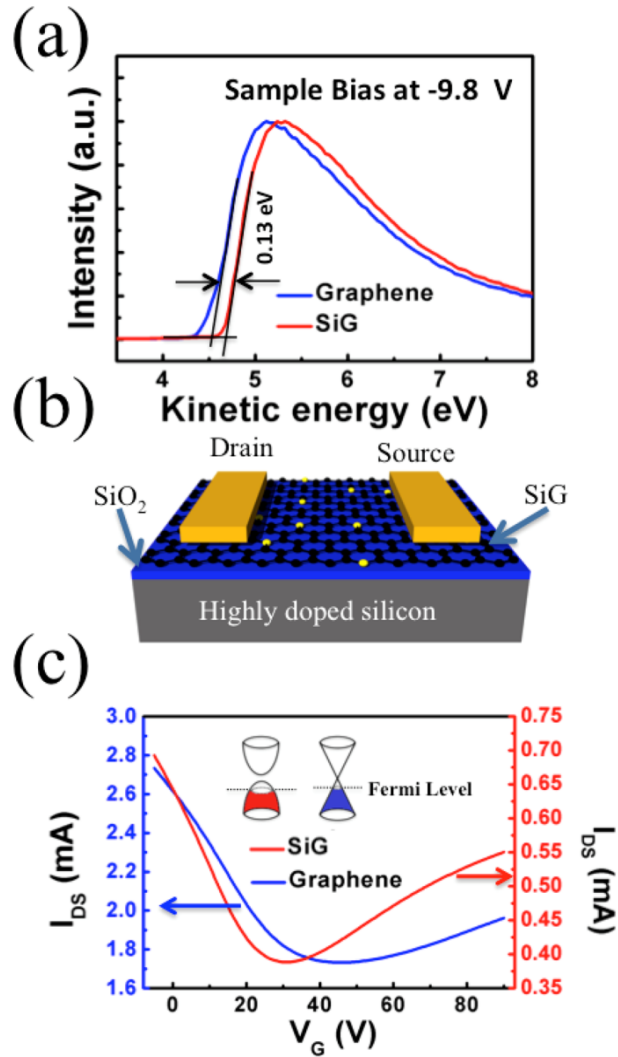

**Fig. 3** (a) UPS of SiG and graphene film. (b) Schematic structure of the FETs based on SiG. (c) Transfer characteristics of SiG and graphene.

In order to evaluate the value of the band gap opened by silicon doping, the first-principle calculations were performed to investigate the electronic structure of SiG. As shown in Fig. 4a, the initial atomic configuration is optimized before band energy calculation, and the percentage of silicon atoms is 3.125% in the SiG sheet. The band structure illustrated in Fig. 4b shows a remarkable band gap at K point with a value of about 0.28 eV.

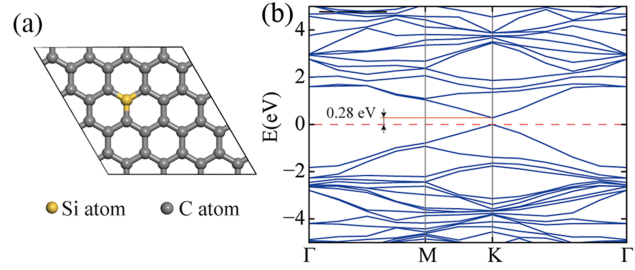

**Fig. 4** (a) Atomic configuration of SiG, the percentage of silicon atoms is 3.125%. (b) Band structure of SiG.

To further explore the potential application of SiG, we used this material to fabricate the SiG/GaAs solar cell. Fig. 5a shows the schematic structure of the solar cell, which comprises silver electrode, SiG membrane, SiNx and GaAs. SiG was transferred to GaAs by wet transferring process. Due to the work function difference between SiG and n-GaAs, the holes in work function difference between SiG and n-GaAs, the holes in GaAs move to SiG and the electrons are left in GaAs, thus forming a Schottky junction with a depletion region in GaAs. When the photons are absorbed in GaAs, the electron-hole pairs are generated and separated by the build-in barrier in the depletion region, leading to the electrons and holes collected by GaAs and SiG, respectively, as shown in Fig. 5b. Fig. 5c displays the current density-voltage (J-V)) curves of the solar cell with SiG and graphene under AM1.5 illumination at 100 mW/cm$^2$. Compared the solar cell using SiG with that of graphene, the short circuit current ($J_{sc}$) and the open circuit voltage ($V_{oc}$) increase from 11.8 mA/cm$^2$ to 13.6 mA/cm$^2$, 0.59 V to 0.65V, with fill factor (FF) of 50% and 51%, respectively, which boost the PCE from 3.5% to 4.5%. According to the measurements of 13 samples of SiG and graphene/GaAs solar cell, a diagram of PCE distribution is illustrated in Fig. 5d, indicating that the PCE of SiG/GaAs solar cell has been generally improved when compared to that of graphene/GaAs solar cell. The average PCE of SiG/GaAs solar cell is 3.69% while that of graphene/GaAs solar cell is 2.76%. Another four specimens of SiG and graphene/GaAs solar cells whose PCE are close to the average value are shown in Fig. S2.† The EQE of the SiG and graphene/GaAs solar are presented in Fig. 5e. From the data, we can find there is a significant improvement of the EQE for the SiG/GaAs solar cell within wavelength range of 300 nm < λ <670 nm, indicating the solar cell with SiG has a better quality at SiG/GaAs interface than the graphene based device.

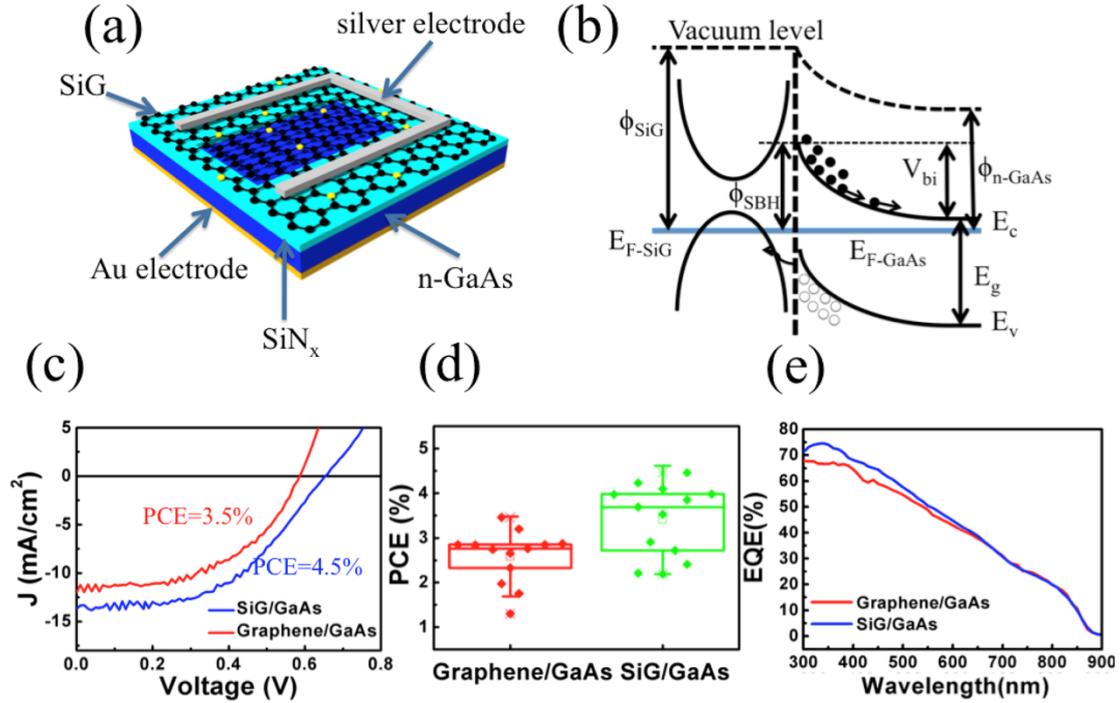

**Fig. 5** (a) Schematic structure of the SiG/GaAs solar cell. (b) The band diagram at the graphene/GaAs interface. (c) J-V curves of SiG/GaAs solar cell and graphene/GaAs solar cell. (d) PCE distribution of SiG and graphene/GaAs solar cell samples. (e) EQE of SiG and graphene/GaAs solar cell.

The enhanced performance of SiG/GaAs solar cell mainly results from the modified electronical properties of graphene by silicon doping. The Schottky barrier height ($\phi_{SBH}$) at the graphene/GaAs interface is determined by the difference of the work function of graphene ($\phi_{gr}$) and the electron affinity of GaAs ($\chi_{GaAs}$), which is expressed as $\varphi_{SBH}=\varphi_{gr}-\chi_{GaAs}$. With the increasing of $\phi_{gr}$, $\phi_{SBH}$ becomes larger. According to the UPS in Fig. 3a, the work function of SiG is 0.13 eV larger than that of graphene, contributing to augment of $\phi_{SBH}$, leading to the improvement of $V_{oc}$. Besides, the dark J-V curve of silicon doped and pristine graphene/GaAs (See Fig. S3†) are described in equation (3):

$$J = J_s\left[\exp\left(\frac{qV}{\eta kT}\right) - 1\right] \quad (3)$$

where J is the current density across the graphene/GaAs interface, q is the value of electron charge, η is the ideality factor, K is the Boltzmann constant, T is the temperature, $J_s$ is the saturation current density expressed by equation (4):

$$J_s = AT^2 \exp\left(-\frac{q\varphi_{SBH}}{kT}\right) \quad (4)$$

where A is the effective Richardson's constant of GaAs (8.9 A/k·cm$^2$).[41] The $J_s$ of graphene/GaAs deduced from eqn (3) is $2.66 \times 10^{-4}$ mA/cm$^2$, and that of SiG/GaAs is $1.98 \times 10^{-5}$ mA/cm$^2$, the Schottky barrier height of graphene/GaAs is 0.74 eV while that of SiG/GaAs is 0.82 eV, the increased $\phi_{SBH}$ contributes to the improvement of $V_{oc}$. In addition, the η of SiG/GaAs deduced from the dark J-V curve is 2.17 and that of graphene/GaAs is 2.38. Due to the high quality interface between SiG and GaAs, the recombination of electrons and holes in the surface of GaAs has been significantly reduced, thus making charge separation more efficient and finally increase the value $J_{sc}$. Analysis of the other SiG and graphene/GaAs solar cells (Fig. S2†) is shown in table S1.†

**Conclusions**

In summary, monolayer SiG in large area has been synthesized using CH$_4$ and SiH$_4$ via CVD method, the strong D Raman band and the clear peak of Si2p in XPS demonstrated the silicon atoms are doped into graphene lattice. Moreover, according to the electrical properties of SiG measured, we can come a conclusion that the band gap of graphene opened by silicon doping can reach 0.28 eV with 3.125% concentration of silicon doped atoms, and the work function of SiG is 0.13 eV larger than pristine graphene. More importantly, compared with the performance of the graphene/GaAs solar cell, the short circuit current and the open circuit voltage of the SiG/GaAs solar cell have a significant improvement, leading to a 33.7% enhancement in average PCE resulting from the increased work function of SiG and a better interface between SiG and GaAs. This improvement of SiG/GaAs solar cell indicates the potential of the SiG in photoelectronic field applications which has not been reported. Our results will cast a new direction on the further research of SiG.

**Experimental**

**Synthesis and transfer of SiG**

Large area SiG synthesis was accomplished via CVD process, the copper foils (99.8% purity, 25 μm thick, Alfa Aesar) were used as substrates for the growth. During the growth process, the reactor was heated up to 1015 °C under Ar and $H_2$ atmosphere, and the copper foils were annealed at the temperature for 30 min, then the mixed flow of $SiH_4$ and $CH_4$ ($SiH_4$ : $CH_4$ = 1 : 50) entered into the furnace to synthesize SiG. Finally, the reactor was quickly cooled down to room temperature. The temperature profile is illustrated in Fig. S4.† The SiG film was transferred to $SiO_2$/Si (300 nm/500 μm), germanium (Ge) or CaAs substrate in following procedures: the SiG on copper foil was coated by PMMA (4000 rpm/min) and baked at 120 °C for 2 min, subsequently, the copper was etched away in $Cu_2SO_4$/HCl/$H_2O$ (16 mg/50 ml/50 ml) solution. Then the SiG film with PMMA was transferred to aimed substrate after cleaning in deionized water. Finally, the PMMA was removed by acetone.

**Characterizations of SiG**

Raman spectra of SiG and pristine graphene membrane were carried out with a Renishaw micro-Raman spectrometer at 532 nm excitation wavelength with 50 ×objective. The ultraviolet photoelectron spectroscopy (UPS) measurements of SiG and pristine graphene were directly taken on graphene above the copper, while the X-ray photoelectron spectroscopy (XPS) was taken on SiG and graphene transferred onto Ge substrate. The suspended SiG membrane was transferred to the carbon transmission electron microscope (TEM) grid for morphology measurement using TEM (Tecnai F-30 operating at 300 KV), while the thickness of SiG was investigated by atomic force microscopy (AFM) (Veeco MultiMode).

**Fabrication and characterization of effect field transistor (FET)**

After cleaning the $SiO_2$/Si (300 nm/500 μm) substrates in the acetone (5 min) and isopropanol (5 min) solvents, the silicon doped and pristine graphene were transferred to the surface of $SiO_2$/Si substrates. Then the In-Ga alloy was pasted onto the surface of silicon to be used as the back electrode, while the Cr/Au (5 nm/60 nm) was thermally deposited on the surface of silicon doped and pristine graphene for the source and drain electrodes respectively. The transfer characteristics were measured by Agilent B1500A system.

**Fabrication and characterization of solar cell**

Cr/Au (5 nm/60 nm) was thermally deposited on GaAs for the back Ohmic contact electrode, 80 nm SiNx was deposited on GaAs by plasma enhanced CVD for dielectric layer with the opened window defined as the solar cell active area. Then the opened window was cleaned in $HCl/H_2O$ (1:3) solution and the SiG or graphene was transferred to the GaAs surface. Finally, the silver electrode was thermally deposited on the SiG or graphene surface above the SiNx dielectric layer. The finished silicon doped and pristine graphene/GaAs solar cells were tested with a solar simulator under AM1.5 illumination at 100 mW/cm$^2$. The current-voltage data were recorded using a Keithley 4200 system. External quantum efficiencies (EQE) of the silicon doped and pristine graphene/GaAs solar cells were measured with PV measurements QEXL system.

**Simulation of the band gap of SiG**

The first-principle calculations were performed using a plane wave open source code QUANTUM-ESPRESSO[42] with the PBE exchange-correlation, Vanderbilt ultrasoft pseudopotentials, and a kinetic energy cutoff energy of 50 Ry.[43] Visualization was performed using VESTA.[44]

**Acknowledgements**

S. S. Lin thanks the support from the National Natural Science Foundation of China (No.51202216, 61431014, 61171037, 61171038, 61322501, 61275183, 61376118, 60990320 and No. 60990322) and Special Foundation of Young Professor of Zhejiang University (No. 2013QNA5007).